 \documentclass[aps,prb,letterpaper,twocolumn,superscriptaddress,showpacs,amsmath,amssymb]{revtex4}

 \usepackage{graphicx}
 \usepackage{dcolumn} 
 \usepackage{bm}      
 \usepackage{verbatim}%
 \usepackage{epstopdf}
 \usepackage[light,timestamp,draft]{draftcopy} 
 \usepackage{hyperref}%

\begin{document}


\title
{
Atomistic simulations of the structures of Pd-Pt bimetallic nanoparticles and nanowires
}


\author{ Kayoung Yun }
\affiliation
{
School of Advanced Materials Engineering, Kookmin University, Seoul 136-702, Korea
}

\author{ Pil-Ryung Cha }
\affiliation
{
School of Advanced Materials Engineering, Kookmin University, Seoul 136-702, Korea
}

\author{ Jaegab Lee }
\affiliation
{
School of Advanced Materials Engineering, Kookmin University, Seoul 136-702, Korea
}

\author{ Jiyoung Kim }
\affiliation
{
Department of Materials Science and Engineering, The University of Texas at Dallas, Richardson, Texas 75080, USA
}

\author{ Ho-Seok Nam }
\email
{hsnam@kookmin.ac.kr}
\affiliation
{
School of Advanced Materials Engineering, Kookmin University, Seoul 136-702, Korea
}
\affiliation
{
Department of Materials Science and Engineering, The University of Texas at Dallas, Richardson, Texas 75080, USA
}

\date{\today}

\begin{abstract}
Bimetallic nanoalloys such as nanoparticles and nanowires are attracting significant attention due to their vast potential applications such as in catalysis and nanoelectronics. Notably, Pd-Pt nanoparticles/nanowires are being widely recognized as catalysts and hydrogen sensors. Compared to unary systems, alloys present more structural complexity with various compositional configurations. Therefore, it is important to understand energetically preferred atomic structures of bimetallic nanoalloys. In this study, we performed a series of simulated annealing Monte Carlo simulations to predict the energetically stable atomic arrangement of Pd-Pt nanoparticles and nanowires as a function of composition based on a set of carefully designed empirical potential models. Both the Pd-Pt nanoparticles and nanowires exhibit quasi-ordered configurations, quite similar to bulk alloy phases such as the L1$_0$ and the L1$_2$ structures with minor surface segregation effects. We believe that this study can provide a theoretical guide for the design of various bimetallic nanomaterials.
\end{abstract}

%


\maketitle


\section{\label{sec:level_1intro} Introduction}

Recently, nanomaterials are drawing increasingly more attention due to their unique physical and chemical properties that are distinct from bulk materials. Among various nano-structured materials, bimetallic nanoparticles and nanowires are receiving the spotlight in many areas such as catalysts, sensors, electronic devices and bio-medicals.~\cite{NanostructuredMaterials} In particular, Pd-Pt bimetallic nanoalloys are well known as hydrogen sensors and catalysts.~\cite{Science2009, JACS2011PdPt} 

Numerous efforts have focused on the synthesis of bimetallic nanoparticles/nanowires,~\cite{jp9704224, ja201156s, JMR:7938489} but their preferred equilibrium structures are not yet well understood. Since the chemical and physical properties of bimetallic nanowires are mainly dependent on their size, compositional configuration and structural stability, it is important to identify their stable atomic structures. 

Compared to unary systems, nanoalloys present more structural complexity because the two components can exhibit various structural modifications. While bimetallic nanoalloys of various structures can be synthesized by a wide variety of techniques, the intrinsic equilibrium structure of bimetallic nanoparticles/nanowires depends on the alloy components. In spite of exhaustive works of both experiment and theory, it is not well established which alloy systems tend to exhibit mixed-alloy or ordered intermetallic nanoalloys. While their structural and phenomenal complexity makes it difficult to experimentally predict the stable atomic structure of bimetallic nanoalloys,~\cite{ja0526618} atomistic simulation could be used as an alternative to understand structural properties. In this paper, we performed a series of simulated annealing (SA) Monte Carlo (MC) simulations based on a set of embedded atom method (EAM) models to investigate energetically preferred atomic structures of Pd-Pt nanoparticles and nanowires.

\section{\label{sec:level_2simul} Computational Details}

\subsection{\label{sec:sublevel_21Potential} Interatomic potentials for Pd-Pt alloys}

One of the key components of atomic scale simulations is the method of evaluating the potential energy of systems and interatomic forces as a function of the positions of atoms. Here, we adopted a set of original and updated versions of the embedded atom method (EAM) potential models for Pd-Pt developed in our previous work.~\cite{Nam:PtBasedNP}

The total energy of the system is given by the usual EAM form:
\begin{equation}
\label{eq:EAM_TotalEnergy}
E ~ = ~ \sum_{i} \left[ F_{s_i} ( \bar \rho_i )
~ + ~ \frac {1} {2} \sum_{j \neq i} \phi_{s_i-s_j}  ( r_{ij} ) \right], 
\end{equation}
where $ F_{s_i} ( \rho )$ is the energy associated with embedding atom of type $s_i$ in a uniform electron gas of density $\rho$ and $\phi_{s_i-s_j} ( r )$ is a pairwise interaction between atoms of type $s_i$ and $s_j$ separated by a distance $r$.  
The electron density is given by
\begin{equation}
\label{eq:EAM_ElectronDensity}
\bar \rho_i = \sum_{j(\ne i)} f ( r_{ij} ), 
\end{equation}
where the atomic electron density function $f(r)$ is taken as the density of a hydrogenic $4s$ orbital following the formalism by A. Voter~\cite{Voter:EAMfccMetals}:

\begin{equation}
\label{eq:EAM_DensityFitting}
f(r) = f_{0}~r^{6} \left( e^{- \beta r} + 2^9 e^{ -2 \beta r} \right).  
\end{equation}
Here, $\beta$ is an adjustable fitting parameter that quantifies the distance over which the electron density decays away from an atom position and $f_0$ is a prefactor. (We chose $f_0$ to be $1/f(r_{eq}N_{1st})$ for convenience.)

The pair potential term $\phi(r)$ is chosen to take a Morse potential form with a minor additional term:
\begin{eqnarray}
\label{eq:EAM_PairPotential}
\phi (r) = 
&& - D_{M} \left[ 2 e^{- \alpha_{M} \left( r - r_{M} \right)} - e^{- 2\alpha_{M} \left( r - r_{M} \right)} \right]
\nonumber\\
&& + \frac{64 \delta}{(r_3 - r_2)^6} (r-r_2)^3 (r_3-r)^3
\nonumber\\
&& \; \; \times \theta (r-r_2) \theta(r_3-r),
\end{eqnarray}
where $D_{M}$, $\alpha_{M}$, $r_{M}$, and $\delta$ are adjustable fitting parameters. Here, the second term in the pair potential function was introduced in order to tune the melting point for unary metals and the heat of mixing for alloys: the parameter $\delta$ represents the magnitude of pair interaction tuning between the second and third nearest neighbor positions, $r_2$ and $r_3$, with the aid of the Heaviside step function $\theta (r)$. 

Finally, the embedding function was numerically determined so that the total energy of reference unary system as a function of dilation satisfies the following universal binding energy relation~\cite{Rose:EAMUniversalFeatures}: 
\begin{equation}
\label{eq:EAM_UniveralRelation}
E(a) = - E_{0} \left[ 1 + \alpha \left( \frac{a}{a_{0}} - 1 \right) \right]
\exp \left[ - \alpha \left( \frac{a}{a_{0}} - 1 \right) \right], 
\end{equation}
with $\alpha = \sqrt{ 9B\Omega / E_0} $, where $a$ is the dilated lattice constant, $a_{0}$ is the equilibrium lattice constant and $E_{0}$, $B$, and $\Omega$ are the cohesive energy, bulk modulus, and equilibrium atomic volume of the reference lattice, respectively. 
The potential interactions were smoothly cut off at $r=r_{cut}$ ($\sim 5.5$ \AA) to ensure that the interatomic potential and its first derivatives are continuous. 

For a binary alloy potential, we need to fit the five adjustable parameters ($D_M, R_M, \alpha_M, \beta$, and $\delta$) for each single component and the four parameters ($D_M, R_M, \alpha_M$, and $\delta$) for a cross-species pair interaction. We adopted the parameter sets for Pd and Pt developed in our previous work,~\cite{Nam:PtBasedNP} but re-optimized the parameters for the Pd-Pt cross-species interaction by including the heat of mixing behavior of the solid and liquid phases and phase diagram information in the target properties. Table~\ref{tab:table1} shows the parameter set for the Pd-Pt system including two different versions of Pd-Pt interactions, which were labelled I and II and hereafter referred as EAM-I and EAM-II, respectively.
\begin{table}[!tbp]
\caption
{\label{tab:table1}
Parameters for the Pd-Pt EAM potentials.
}
\begin{ruledtabular}
\begin{tabular}{cccccc}
 & $\beta$ (\AA$^{-1}$)   & $D_M$ (eV)  & $\alpha_M$ (\AA$^{-1}$) & $r_M$ (\AA) & $\delta$ (eV) \\
\hline \\  [-1.5ex]

Pd         & 3.400 & 1.682 & 1.501 & 2.343 & 0.06  \\  [0.5ex]
Pt         & 3.580 & 0.783 & 1.779 & 2.577 & 0.01  \\  [0.5ex]
Pd-Pt (I)  &   -   & 1.259 & 1.701 & 2.448 & 0.00  \\  [0.5ex]
Pd-Pt (II) &   -   & 1.254 & 1.696 & 2.443 & 0.06  \\  [0.5ex]
\end{tabular}
\end{ruledtabular}
\end{table}

\subsection{\label{sec:sublevel_22Properties} Structural properties of Pd-Pt bulk alloys}

Compared to unary systems, alloy systems present more structural complexity due to various compositional configurations. Since atomic structures of nanomaterials are generally determined by competition between bulk mixing behavior and surface effects, the bulk alloy structure and surface energy seem to be two key factors to be considered. However, in spite of being an apparently simple binary system, only a few old articles discuss its thermodynamic properties and bulk alloy phase behavior of the Pd-Pt system.~\cite{Darby:ThermoPdPt, Bharadwaj1991167}

\begin{figure}[!tbp]
\centering
\includegraphics[width=0.45\textwidth]{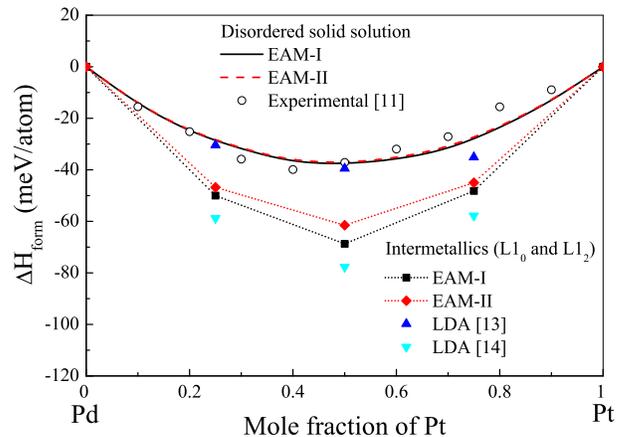}
\caption
{ \label{fig:dHmix}
Heat of formation for the Pd-Pt bulk system as a function of Pt concentration: The solid (black) and dashed (red) lines are the results of the present EAM models for a disordered solid solution (random mixing) while the open circles are those of experimental measurement (Ref.~\onlinecite{Darby:ThermoPdPt}). The formation energies of three possible intermetallics (L1$_2$-$\rm Pd_{3}Pt$, L1$_0$-PdPt, and L1$_2$-$\rm PdPt_{3}$) are also represented by solid symbols: The squares and diamonds with dotted lines correspond to the results of the present EAM models and the other solid symbols to first-principle calculation data from Refs.~\onlinecite{PhysRevLett.66.1753} and \onlinecite{PhysRevB.74.064202}. 
}
\end{figure}

Figure~\ref{fig:dHmix} shows the calculated heats of formation for a disordered solid solution (random mixing) and some possible intermetallics in comparison with experimental and first-principle calculation data on the Pd-Pt system in the literature.~\cite{Darby:ThermoPdPt, PhysRevLett.66.1753, PhysRevB.74.064202} At higher temperatures, the Pd-Pt system forms a completely miscible solid solution based on the face-centered cubic (fcc) lattice~\cite{Bharadwaj1991167} and both the EAM-I and EAM-II potentials agree well with the experimental data. However, the Pd-Pt system has not been studied in  detail at low temperatures, and there has been controversy over whether or not any stable intermetallic compounds exist. For example, most experiments and some first-principle calculations based on local density approximation (LDA) propose that L1$_0$ is the most stable compound for Pd-Pt of 1:1 stoichiometry,~\cite{PhysRevLett.66.1753, PhysRevB.74.064202} while some other LDA calculations predict that L1$_1$ structure has the lowest energy.~\cite{Curtarolo2005163, PhysRevX.3.041035} 

Table~\ref{tab:table2} provides the calculated formation energies of some fcc based alloy crystal structures in comparison with those by first-principle calculations.~\cite{PhysRevLett.66.1753, PhysRevB.74.064202} As shown in Fig.~\ref{fig:dHmix}, both the EAM potentials give reasonable values for the formation energy of L1$_0$ and L1$_2$ intermetallic compounds falling in the middle of the range of Refs.~\onlinecite{PhysRevLett.66.1753} and \onlinecite{PhysRevB.74.064202}. As indicated in Table~\ref{tab:table2}, the formation energies of the D0$_{22}$ ($\rm Pd_{3}Pt$ and $\rm PdPt_{3}$) structures are lower than that of L1$_2$ structure (by less than 1 meV) in the case of the EAM-I potential, while the EAM-II model predicts the $(100)$ family of ordered states that consists of L1$_2$ for $\rm Pd_{3}Pt$ and $\rm PdPt_{3}$ and L1$_0$ for PdPt in accordance with Refs.~\onlinecite{PhysRevLett.66.1753} and \onlinecite{PhysRevB.74.064202}.
\begin{table}[!tbp]
\caption
{\label{tab:table2}
Calculated enthalpies of formation for the possible $\rm Pd_{1- \it x} \rm Pt_{\it x}$ intermetallics: the formation energies of some face-centered cubic (fcc) based alloy crystal structures (L1$_0, L1_1, L1_2, D0_{22}$) are given in comparison with those by first-principle calculations. The numbers with * indicate the formation energies of the lowest energy structure.
}
\begin{ruledtabular}
\begin{tabular}{ccccc}
Crystal Structure & EAM-I & EAM-II & LDA~\cite{PhysRevLett.66.1753} & LDA~\cite{PhysRevB.74.064202} \\
\hline \\  [-1.5ex]
L1$_0    (PdPt)$      & -68.7*  & -64.2*  & -39.5*  & -77.7* \\
L1$_1    (PdPt)$      & -24.9   & -26.3   & -29.1   & -68.4  \\
L1$_2    (Pd_{3}Pt)$  & -50.0   & -46.8*  & -30.4*  & -58.8* \\
D0$_{22} (Pd_{3}Pt)$  & -50.6*  & -44.0   & -22.1   & -54.4  \\
L1$_2    (PdPt_{3})$  & -48.2   & -45.1*  & -35.1*  & -57.8* \\
D0$_{22} (PdPt_{3})$  & -48.7*  & -41.9   & -29.1   & -51.3  \\
\end{tabular}
\end{ruledtabular}
\end{table}

\begin{figure}[!tbp]
\includegraphics[width=0.45\textwidth]{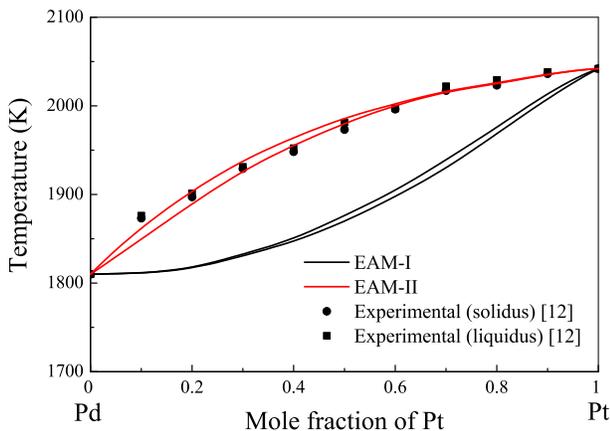}
\caption
{ \label{fig:PhaseDiagram}
Calculated phase diagram of the Pd-Pt binary alloy (solid lines) in comparison with experimental data (solid symbols, Ref.~\onlinecite{Bharadwaj1991167}).
}
\end{figure}

We also calculated the solid-liquid phase diagrams for the Pd-Pt bulk alloys using Gibbs-Duhem integration technique combined with semigrand canonical Monte Carlo simulations.~\cite{Kofke:GibbsDuhemMolPhys, PhysRevB.75.014204} Figure~\ref{fig:PhaseDiagram} shows calculated phase diagrams of the Pd-Pt binary alloy in comparison with experimental data.~\cite{Bharadwaj1991167} Although the parameter set of the EAM-II potential was marginally modified from that of the EAM-I potential, it produced results almost identical to those produced by the EAM-I potential in predicting the enthalpy of formation for a disordered solid solution (random mixing). However, their solid-liquid phase diagrams are quite different: Since the potential parameters are adjusted to reproduce the heats of formation for both the solid and liquid phases, the solid-liquid phase diagram predicted by the EAM-II model agrees much better with the experiments. 

In nearly all of the previous simulation studies on Pd-Pt nanoalloys,~\cite{PhysRevB.74.155441, PhysRevB.76.134117, PdPtMDMC, B207847C, B313811A, GAPdPt, Cheng:Onion2006, Cheng:Onion2008} a simplified combination of the Pt-Pt and Pd-Pd parameters was set as potential parameters for the cross-species interaction instead of fitting to Pd-Pt alloy properties. Consequently, bulk Pt-Pd alloys were assumed to be solid solutions, rather than ordered intermetallics, for all compositions.~\cite{PhysRevB.74.155441, PhysRevB.76.134117, PdPtMDMC, B207847C, B313811A, GAPdPt, Cheng:Onion2006, Cheng:Onion2008} However, our EAM models for Pd-Pt alloys, which were fitted to alloy properties such as the formation enthalpy and the lattice constant, rather predict stable intermetallic compounds in accordance with the first-principle calculations. 

Although the bonding characteristics of materials originate from their electronic structures, in classical atomistic simulations, such as molecular dynamics (MD) and Monte Carlo methods, the interactions between atoms are usually described by empirical potentials, and therefore the reliability of the simulation results is entirely dependent on the reality and accuracy of the interatomic potentials. So far, lack of appropriate potential models for bimetallic systems have attributed to insufficiencies of atomic scale simulations on the structural properties of alloy nanoalloys. Considering that the present EAM potentials for the Pd-Pt alloy system well reproduce the formation energies of the alloy phases as well as the experimental phase diagram, the present empirical model approach may provide reliable simulations of alloy nanostructures such as nanoparticles and nanowires.

\subsection{\label{sec:sublevel_23Simulation} Monte Carlo simulations of nanoalloys }

\begin{figure}[!tbp]
\includegraphics[width=0.48\textwidth]{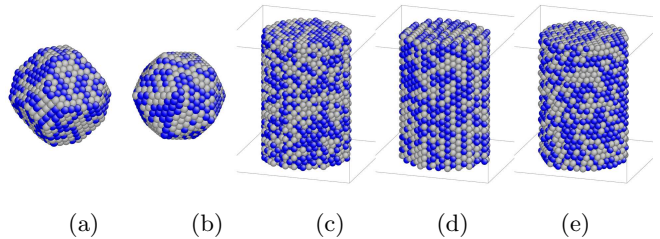}
(a) \hspace{1.00cm} (b) \hspace{1.00cm} (c) \hspace{1.00cm} (d) \hspace{1.00cm} (e)
\caption
{ \label{fig:initialcfg}
Initial atomic configurations: bimetallic nanoparticles of random mixing with (a) truncated octahedron (TOh) and (b) icosahedron (Ih) shapes and bimetallic nanowires of random mixing with (c) [100], (d) [110], and (e) [111] crystallographic orientations along the nanowire axis.
}
\end{figure}

It is well known that metallic nanoparticles present competitive structural motifs, such as the icosahedron(Ih), decahedron(Dh), cuboctahedron, and truncated octahedron (TOh).~\cite{Nam:IcosahedronPRL, Nam:IcosahedronPRB} In our atomistic simulations of alloy nanoparticles, truncated octahedral (TOh) and icosahedral (Ih) nanoparticle of $\sim$ 4 nm in diameter (1654 and 1415 atoms, respectively) were used as initial atomic configurations. Initial compositional configurations were typically generated by random mixing of atoms as shown in Figs.~\ref{fig:initialcfg}(a) and (b). 

For the simulations of nanowires, cylindrical shape nanowires of $\sim$ 4 nm in diameter (also with random mixing) were used as initial structures. The bimetallic nanowires were chosen to have single-crystalline fcc structures with three different crystallographic orientations, i.e., [100], [110], and [111] along the nanowire axis, as shown in Figs.~\ref{fig:initialcfg}(c)-(e). We also investigated the bimetallic nanowire structure of various mole fractions: the composition of 25, 50 and 75 at\% Pt. The nanowires were modelled as infinitely long wires by applying the periodic boundary condition along the wire axis.

In order to investigate energetically preferred atomic structure of Pd-Pt nanoalloys, we carried out a series of simulated annealing  Monte-Carlo optimizations based on Metropolis algorithm.~\cite{Frenkel:UnderstandingMS} The simulated annealing procedure was performed from $T =$ 1000 K to 0 K at a cooling rate of $\sim$ 1 K/MCS (where MCS stands for Monte Carlo step) in order to avoid being isolated in a local minimum. In all of the MC simulations, lattice relaxation using the conjugate-gradient method was implemented after configurational exchange of two select atoms.

\section{\label{sec:level_3Result} Results and discussion}

\subsection{\label{sec:sublevel_31Nanoparticles} Structure of Pd-Pt bimetallic nanoparticles }

We found that the optimized structures were relatively insensitive to the initial structure or cooling rate of the simulated annealing procedure, although there were some minor changes in surface atomic structure due to the random nature of Monte Carlo simulations. The consistent results may be attributed to the low transition barrier for compositional configuration change from a random structure to the optimized structure in the simulations. 

\begin{figure}[!tbp]
\includegraphics[width=0.48\textwidth]{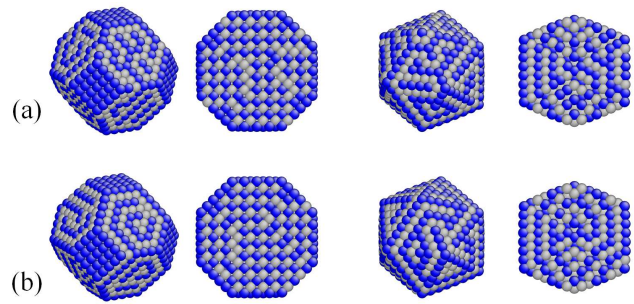}
\caption
{ \label{fig:NPsnapshot}
Predicted atomic structures of Pd-Pt nanoparticles simulated with (a) the EAM-I and (b) the EAM-II potentials:
The first and second columns of the figures show each of their surface and cross-sectional views of the truncated octahedron (TOh) nanoparticles while the third and fourth columns of the figures show each of their surface and cross-sectional views of the icosahedron (Ih) nanoparticles, respectively. Pd atoms are in blue (dark) and Pt in grey (light).  
}
\end{figure}

Figure~\ref{fig:NPsnapshot} shows the typical optimized atomic structures of Pd-Pt nanoparticles of 1:1 stoichiometry that were simulated with the EAM-I and the EAM-II potentials. Both the EAM potential models produced similar inner structures with some minor difference in surface atomic structure: The cross-sectional views of truncated octahedron Pd-Pt nanoparticle depict a typical L1$_0$ intermetallic compound structure with surface segregation of Pd, while alloy nanoparticles of icosahedral shape exhibit somewhat complicated patterns in their cross-sectional views, probably due to the intrinsic twin boundaries of the icosahedral nanoparticles. Note that all of the $\{100\}$ type facets of TOh nanoparticles are occupied only by Pd atoms, while $\{111\}$ type facets are composed of both Pd and Pt atoms with linearly aligned patterns. In both the TOh and Ih nanoparticles, edges are mostly occupied by Pd atoms.
\begin{figure}[!tbp]
\includegraphics[width=0.45\textwidth]{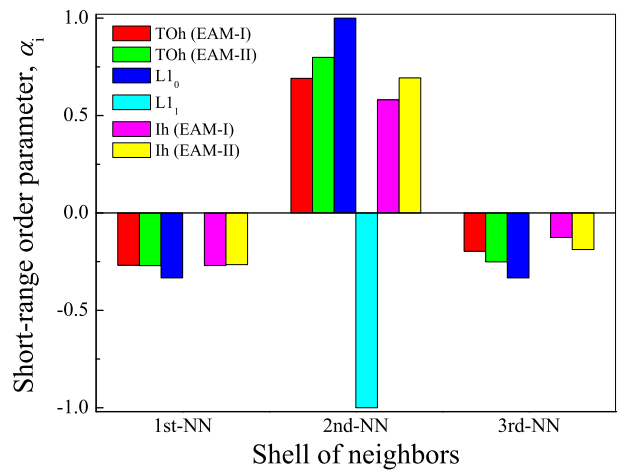}
\caption
{ \label{fig:NPsrop}
Short-range order parameter $\alpha_i$ for the TOh and Ih nanoparticles in comparison with those of the ideal ordered alloy structures. The order parameter $\alpha_i$ was calculated for the first three shells of nearest-neighbors(NN). 
}
\end{figure}

In order to clearly understand the inner structures, the bond characteristics of the alloy nanoparticles were analyzed by using the short-range-order parameter proposed by Cowley~\cite{Cowley:CSROPRB} and Warren.~\cite{Warren:CSROJAP} The Warren-Cowley short-range order parameter $\alpha_i$ is defined as
\begin{equation}
\label{eq:alphaSROP}
\alpha_i = 1-p_{i,AB}/C_B = 1- \frac {N_{i,AB}} {N_{i} \cdot C_B},
\end{equation}
where $p_{i,AB}$(=$N_{i,AB}$/$N_{i}$) is the probability that the atomic site is occupied by a $B$ atom in the $i$th-neighbor shell of an $A$ atom, $N_{i,AB}$ is the expected number of $B$ atoms around an $A$ atom in the $i$th shell of neighbors, $N_{i}$ is the total neighbor number in the $i$th-neighbor shell, and $C_B$ is the atomic concentration of $B$. All of the $\alpha_i$ values range between -1 to 1 and positive values denote more $A$-$A$ and/or $B$-$B$ pairs and negative values denote more $A$-$B$ pairs between the central and the neighbors in the $i$th shell, compared to those of random mixing. The parameter for the first-nearest neighbor shell, $\alpha_1$, is often referred as chemical short-range order parameter and is a useful measure of the chemical affinity representing the degree of tendency for ordering: positive $\alpha_1$ indicates clustering or phase separation and negative $\alpha_1$ indicates strong $A$-$B$ bond and chemical ordering.

Figure~\ref{fig:NPsrop} shows the first three short-range order parameters $\alpha_i$ for the TOh and Ih nanoparticles. (Here, $\alpha_i$ were averaged over several configurations obtained in simulations performed with different random number seeds.) The $\alpha_i$ for the ideal L1$_0$ and L1$_1$ ordered alloy structures were also given for comparison: For example, for the ideal L1$_1$ lattice, $\alpha_{2}=-1$, while $\alpha_1$ and $\alpha_3$ are zero. For the L1$_0$ alloy structure, $\alpha_{2}=1$, while $\alpha_1$ and $\alpha_3$ are 1/3. On the other hand, in the case of randomly mixed alloys (not shown here), $\alpha_i$ tend to zero for any $i$-th neighbor shells. 

As indicated by the distributions of $\alpha_i$ shown in Fig.~\ref{fig:NPsrop}, atomic structures of the TOh Pd-Pt nanoparticles simulated with both the EAM potentials exhibit very similar bond characteristics to that of the L1$_0$ alloys: both $\alpha_1$ and $\alpha_3$ are negative toward -1/3 and $\alpha_2$ is strongly positive. The distributions of $\alpha_i$ for the Ih nanoparticles also show similar patterns as that of the TOh nanoparticles (indicating multiply-twinned L1$_0$ structures), although the absolute value of $\alpha_i$ is relatively small. Interestingly, while the values of $\alpha_1$ are almost identical in both the EAM models, $\alpha_2$ and $\alpha_3$ by the EAM-II model are closer to 1/3, the ideal value for the L1$_0$ structure, which indicates that the nanoparticle structures predicted by the EAM-II model are more ordered with longer range-order.

\subsection{\label{sec:sublevel_32Nanowires} Structure of Pd-Pt bimetallic nanowires }

In the case of the Pd-Pt nanowires, both the EAM-I and EAM-II potentials predicted nearly identical views of the structures. Therefore, we limit our focus to the simulations based on the EAM-II potentials (of more accurate phase diagram behavior) in the remainder of this paper. 

\begin{figure}[t]
\includegraphics[width=0.48\textwidth]{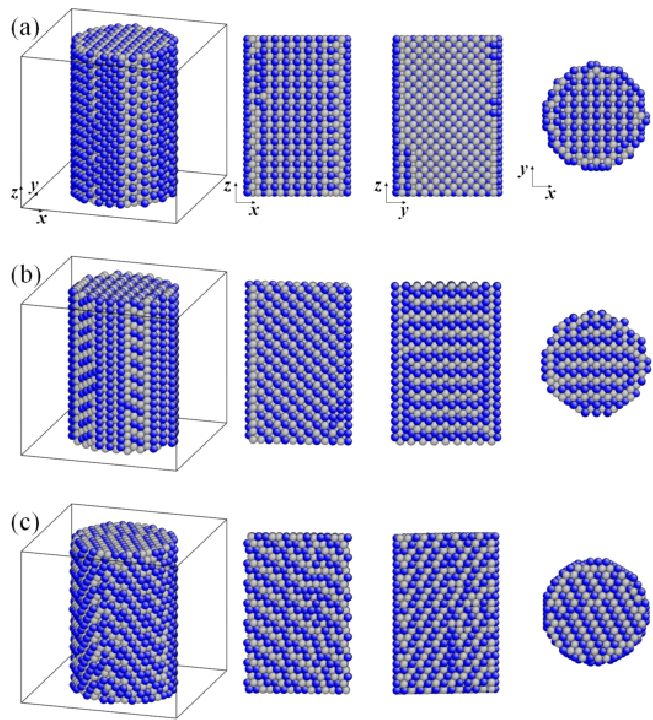}
\caption
{ \label{fig:NWsnapshot1}
Predicted atomic structures of Pd-Pt nanowires with (a) [100], (b) [110], and (c) [111] crystallographic orientations along the nanowire axis. The first column of the figures shows the tilt view of the nanowires, while the second, third, and last columns of the figures show cross-sectional views of the $xz$-, $yz$-, and $xy$- planes, respectively. Pd atoms are in blue (dark) and Pt in grey (light).
}
\end{figure}

Figure~\ref{fig:NWsnapshot1} shows the typical optimized atomic structures of Pd-Pt nanowires with three different crystallographic orientations, i.e., [100], [110] and [111] along the nanowire axis. In all three cases, the inner patterns of alternating distribution of Pd and Pt atomic planes appear to be a minor modification of the L1$_0$ alloy structure with Pd atoms segregated on the surface. Interestingly, the alternating $\{100\}$ type planes of Pd and Pt atoms are mostly aligned parallel to the nanowire axis in the case of [100] oriented nanowires probably because it reduces the surface energy and internal stresses. In the [110] and [111] oriented nanowires, alternating (100) planes of Pd and Pt are inclined according to their crystal orientation angles.
\begin{figure}[!t]
\includegraphics[width=0.48\textwidth]{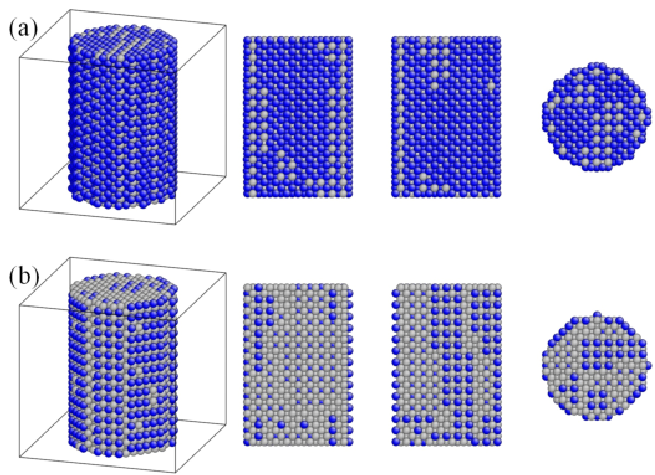}
\caption
{ \label{fig:NWsnapshot2}
Predicted atomic structures of Pd-Pt nanoparticles with compositions of (a) 25\%, (b) 75\% Pt. The first column of the figures shows the tilt view of the nanowires, while the second, third, and last columns of the figures show cross-sectional views of the $xz$-, $yz$-, and $xy$- planes, respectively. Pd atoms are in blue (dark) and Pt in grey (light).
}
\end{figure}

Figures~\ref{fig:NWsnapshot2}(a) and (b) show the typical optimized atomic structures of Pd-Pt nanowires with compositions of 25 and 75\% of Pt, respectively. In the case of 3:1 or 1:3 compositions, the internal structures of nanowires show that minor element atoms occupied the lattice sites with regular patterns. Although the minor element sites do not show perfect long-range order, quasi-ordered atomic arrangement appears to be that of the L1$_2$ bulk alloy phases, as shown in the cross-sectional views of Fig.~\ref{fig:NWsnapshot2}. Compared to the distribution of Pt atoms on the surface of 25\% Pt nanowires, Pd atoms are strongly segregated to the surface for the nanowires of 75\% Pt presumably due to the lower surface energy of Pd. On the other hand, Pt atoms tend to be segregated in the layers beneath the surface (instead of the surface layers) at its low concentration, as shown in Fig.~\ref{fig:NWsnapshot2}(a)

\begin{figure}[!tbp]
\includegraphics[width=0.45\textwidth]{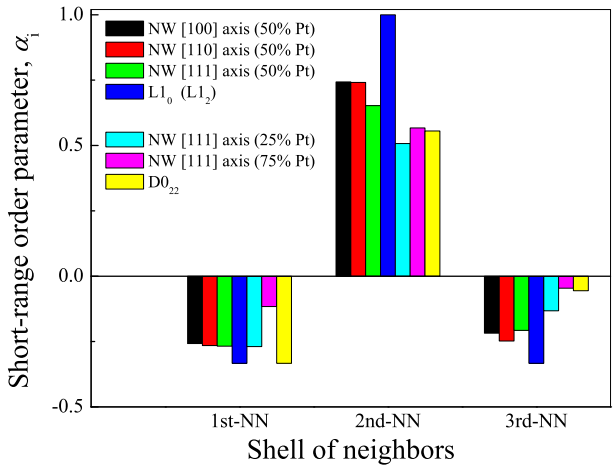}
\caption
{ \label{fig:NWsrop}
Short-range order parameter $\alpha_i$ for the nanowires in comparison with those of the ideal ordered alloy structures. The order parameter $\alpha_i$ was calculated for the first three shells of nearest-neighbors(NN). 
}
\end{figure}

Figure~\ref{fig:NWsrop} shows the short-range order parameters $\alpha_i$ calculated for the optimized structures of the nanowires. The $\alpha_i$ for the ideal L1$_2$ (same as L1$_0$) and D0$_{22}$ ordered alloy structures were also given for comparison. As indicated by the distributions of $\alpha_i$ shown in Fig.~\ref{fig:NWsrop}, all of the Pd-Pt nanowires of 50\% Pt exhibit very similar bond characteristics to that of the L1$_0$ alloys with $\alpha_1<$-0.25, $\alpha_3<$-0.2, and $\alpha_2>$0.2. The distributions of $\alpha_i$ for the 25\% and 75\% Pt nanowires also show similar patterns as those of the ordered structures, although it cannot be absolutely identified as either the L1$_2$ or the D0$_{22}$ structure. Considering that their inner lattice does not satisfy 3:1 or 1:3 compositions due to surface segregation effects, their internal structures are more like those of ordered structures at various diameters up to several nanometers.

Generally, equilibrium compositional structure of bimetallic nanoalloys may be either similar to the bulk alloys or different from the bulk by the competition between atomic mixing behavior and surface effects. Table~\ref{tab:table3} shows the calculated surface energies for various surface types in the Pd-Pt system. Although the $(100)$ surface energy of Pd is substantially lower than that of Pt, the $(111)$ surface energy of Pd is comparable to that of Pt. The surface energy values of randomly-mixed alloys are between those of pure Pd and Pt and the surface energies of the L1$_0$ alloy are even lower than those of disordered alloys. Because of the comparable surface energy properties between those surfaces, the Pd-Pt nanoalloys may maintain a similar compositional structure to the bulk ordered phases without dramatic surface modification, even at this nanoscale regime.

\begin{table}[!tbp]
\caption
{\label{tab:table3}
Calculated surface energies for various surface types in the Pd-Pt system.
}
\begin{ruledtabular}
\begin{tabular}{ccccc}
Bulk structure  & \multicolumn{2}{c}{Surface energy [mJ/m$^2$]} \\
                & \multicolumn{1}{c}{(100)} & \multicolumn{1}{c}{(111)} \\
\hline \\  [-1.5ex]
Pd (fcc)          & 1815  & 1700   \\
Pt (fcc)          & 1891  & 1676   \\
PdPt (disordered) & 1847  & 1678   \\
PdPt (L1$_{0}$)   & 1755  & 1590   \\
\end{tabular}
\end{ruledtabular}
\end{table}

\section{\label{sec:level_4Summary} Summary}
In experiments, energetics is not the only factor that affects the morphology of nanoalloys such  as nanoparticles and nanowires.
In cases where the formation is determined by kinetics rather than thermodynamic factors, the details of individual synthesis processes can govern the structures of the synthesized nanoalloys. Nevertheless, identifying energetically stable compositional configurations may be a first step towards understanding structural stability.

For detailed structural information of the energetically stable bimetallic nanoparticles and nanowires, we performed a series of Monte Carlo optimizations with simulated annealing procedure based on a set of carefully designed EAM potential models. The EAM-I and EAM-II models predict fairly similar structures for nanoalloys, although they are distinguishable in their solid-liquid phase diagram behavior. Our simulations also investigated the energetically favored atomic structures of Pd-Pt nanoparticles and nanowires with various shapes and compositions. Both Pd-Pt nanoparticles and nanowires exhibit quasi-ordered configurations, quite similar to bulk alloy phases such as the L1$_0$ and the L1$_2$ structures with minor surface segregation effects. 

Our simulations quantify the energetically favorable atomic arrangements of the Pd-Pt nanoalloys and explain why they exhibit such equilibrium structures in terms of their bulk alloy structures and surface energies. We believe that our simulation results may be useful reference for further investigation of Pd-Pt nanoalloys. Above all, we expect that this study will provide a theoretical standard for design of bimetallic nanoparticles and nanowires with various possible structures.


\begin{acknowledgements}
This work was supported by Leading Foreign Research Institute Recruitment Program (2013K1A4A3055679) and the Priority Research Centers Program (2009-0093814) through the National Research Foundation of Korea (NRF) funded by the Ministry of Science, ICT \& Future Planning (MSIP).
\end{acknowledgements}


\bibliography{PdPt_Nanowire_arXiv} 

\end{document}